# IS THE HALO RESPONSIBLE FOR THE MICROLENSING EVENTS?[1]

E. ROULET, S. MOLLERACH and G. F. GIUDICE
*Theory Division, CERN*
*Geneva, CH-1211, Switzerland*

ABSTRACT

We discuss whether the astrophysical objects responsible for the recently reported microlensing events of sources in the Large Magellanic Cloud can be identified as the brown dwarf components of the spheroid of our galaxy, rather than the constituents of a dark baryonic halo.

---

[1] Talk given by E. R. at the $6^{th}$ Workshop on Neutrino Telescopes, Venice, february 1994

The microlensing events in the Large Magellanic Cloud (LMC) recently reported by the EROS [1] and MACHO [2] collaborations indicate the presence of compact dark objects passing very close to the line of sight to the monitored stars [3–5]. These objects have been interpreted as the constituents of the dark halo of the Galaxy. We have recently proposed an alternative interpretation [6] in which the dark lensing objects are, instead, the main constituents of the galactic spheroid. This interpretation is suggested by models of the Galaxy based on dynamical observations [7–9], which predict a spheroid component much heavier than accounted for by direct measurements of star counts at high latitudes and of high-velocity stars [10], namely $M_S(\text{lum}) = 1\text{--}3 \times 10^9 M_\odot$, while $M_S(\text{dyn}) = 5\text{--}7 \times 10^{10} M_\odot$. The difference between the luminous and the dynamical determinations of the spheroid mass can be understood if most of the mass in the spheroid is non-luminous, in the form of very faint stars or brown dwarfs, that are then ideal candidates for lensing objects. This is supported by the measured shape of the mass function of spheroid field stars, which behaves, below $m \simeq 0.4 M_\odot$ and down to the smaller masses observed ($0.14 M_\odot$), as $dN/dm \propto m^{-4.5}$ [11], suggesting that most of the spheroid mass is in stars lighter than the hydrogen burning limit $m_{HB} \simeq 0.085 M_\odot$ and hence dark. In our proposed scenario, the predicted event rate for sources in the LMC results lower than in the halo one, but it is still significant and consistent with the observations (that in fact seem to be in deficit with respect to the predictions of the halo model). Yet another possibility has been pointed out by Gould [12], who suggested that the lensing objects could be in the thick disk of the Galaxy.

Spiral galaxies, as our own, have at least three main components: the exponential disk, containing most of the observed stars (that are young and metal-rich) and gas clouds; the halo, which should account for the flatness of the rotation curves and provides most of the mass of the galaxies, but contains essentially no luminous matter; and the spheroid component, a roughly spherical distribution of old (metal-poor) stars, which mass density in our galaxy decreases as $r^{-3.5}$. As mentioned before, the spheroid mass of the Galaxy inferred from dynamical observations is comparable to the total mass of the disk, and is an order of magnitude larger than the mass contained in the observed spheroid stars, indicating a 'missing mass' problem in the spheroid. One could also mention that the relative importance of the disk and spheroid luminosities is one of the main characteristics to identify the Hubble type of a spiral galaxy, with Sc having very small spheroids while Sa have large ones (and ellipticals being 'only spheroid' galaxies).

We use for the different components the following simplified mass densities, which are good approximations to those obtained using fits to dynamical observations:

$$\rho^H = \rho_0^H \frac{r_0^2 + a^2}{r^2 + a^2}, \quad (1)$$

for the halo, and

$$\rho^S = \rho_0^S \left(\frac{\sqrt{r_0} + \sqrt{b}}{\sqrt{r} + \sqrt{b}}\right)^7, \quad (2)$$

for the spheroid, where $r_0 = 8.5$ kpc is the distance from the sun to the galactic centre, $a = 3$ kpc, $b = 0.17$ kpc and the local densities are $\rho_0^H = 0.01 M_\odot/\text{pc}^3$ and $\rho_0^S = 0.0015 M_\odot/\text{pc}^3$ (for reference, the local density of disk stars is approximately $0.1 M_\odot/\text{pc}^3$).

The light curve of a microlensing event is an achromatic and symmetric amplification of the source star light given by

$$A(u) = \frac{u^2 + 2}{u\sqrt{u^2 + 4}}, \quad (3)$$

where $u$ is the distance of the lens to the line of sight in units of the Einstein radius $R_E$

$$u^2(t) = \left(\frac{d}{R_E}\right)^2 + \left(\frac{v_{r\perp}(t - t_o)}{R_E}\right)^2, \quad (4)$$

with $d$ the impact parameter, $v_{r\perp}$ the lens velocity relative to the microlensing tube (see below) in the direction orthogonal to the line of sight, and

$$R_E \equiv 2\sqrt{\frac{GM(L-x)x}{c^2 L}}, \quad (5)$$

with $L$ the distance to the source and $x$ that to the lens of mass $M$. The maximum amplification is reached at the time $t_o$, when $u$ takes its minimum value $u_{min} = d/R_E$. The light curve can then be fitted by the three parameters $t_o$, $A_{max} = A(u_{min})$ and the event duration $T \equiv R_E/v_{r\perp}$.

The probability that a star is being microlensed at a given time is the so-called optical depth $\tau$

$$\tau = \int_0^L dx \frac{d\tau}{dx}, \quad (6)$$

Table 1: The optical depth ($\tau$), the average distance to the lenses ($\langle x \rangle$), the event rate ($\Gamma$), and the average event duration ($\langle T \rangle$), for microlensing events in the LMC. The lensing object mass $m_{0.05}$ is in units of $0.05 M_\odot$. We have taken $u_T = 1$ and $\tau \propto u_T^2, \Gamma \propto u_T$.

| Model | $\tau$ | $\langle x \rangle$ [kpc] | $\Gamma$ [$\frac{\text{events}}{\text{yr } 10^6 \text{stars}}$] | $\langle T \rangle$ [days] |
|---|---|---|---|---|
| Halo | $5.8 \times 10^{-7}$ | 16 | $9.2/\sqrt{m_{0.05}}$ | $15\sqrt{m_{0.05}}$ |
| Spheroid | $4.6 \times 10^{-8}$ | 11 | $0.7/\sqrt{m_{0.05}}$ | $16\sqrt{m_{0.05}}$ |

with
$$\frac{d\tau}{dx} = \frac{4\pi G u_T^2 x(L-x)\rho(x)}{c^2 L}. \qquad (7)$$
A special property of $\tau$ is that it is independent of the mass function of the lensing objects $dN/dm$. The first column of Table 1 shows the value of $\tau$ for the halo and spheroid models.

In Fig. 1, we plot the function $(L/\tau)d\tau/dx$ that measures the contribution of the lenses to $\tau$ in terms of their distance to us. Defining
$$\langle x \rangle \equiv \frac{1}{\tau} \int_0^L dx\, x \frac{d\tau}{dx}, \qquad (8)$$
that is the average distance at which microlensing takes place, we obtain $\langle x \rangle = 11$ kpc and 16 kpc for the spheroid and halo models, respectively (as shown in Table 1). This shows that although the spheroid density decreases very steeply when moving out of the galactic centre, $\langle x \rangle$ is still relatively large for the spheroid, partly because the LMC line of sight is at nearly $80°$ from the galactic centre; hence it is only for points along the line of sight at $x > 10$ kpc that the galactocentric distance starts to increase significantly (we show ticks corresponding to galactocentric distances $r = 10, 20, 30$ and 40 kpc in Fig. 1).

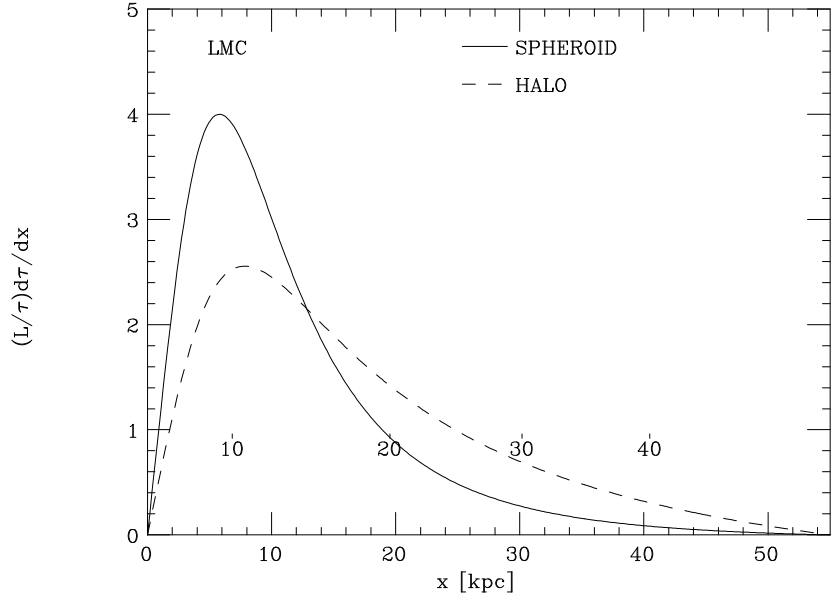

Fig. 1: Contribution to $\tau$ vs. the distance to the lensing objects $x$ for the LMC.

More important than $\tau$ is the microlensing rate $\Gamma$. Assuming an isotropic and isothermal Maxwellian velocity distribution for the lenses, with one-dimensional dispersion $\sigma$, the differential rate is given by

$$\frac{d\Gamma}{dT} = \int_0^\infty dm \int_0^L dx \frac{dN(x)}{dm} 8 u_T \sigma^2 z^4 e^{-(z^2+\eta^2)} I_0(2\eta z), \qquad (9)$$

where $z \equiv R_E/(\sqrt{2}\sigma T)$ and $\eta \equiv |(x/L)\vec{v}_{s\perp} + (1 - x/L)\vec{v}_{\odot\perp}|/(\sqrt{2}\sigma)$, with $\vec{v}_{s\perp}$ and $\vec{v}_{\odot\perp}$ the component of the source and the sun velocity in the direction orthogonal to the line of sight. If the total gravitational potential of the Galaxy leads to a constant rotation velocity $v_c$, an isothermal component with $\rho \propto r^{-n}$ should have a velocity dispersion $\sigma = v_c/\sqrt{n}$; hence, for the spheroid we get $\sigma^S \simeq 120$ km/s, in agreement with the observations, and for the halo $\sigma^H \simeq 155$ km/s. The total event rate, $\Gamma \equiv \int_0^\infty dT (d\Gamma/dT)$, is shown in column 3 of Table 1, assuming that all the lenses have the same mass $M$, i.e. $dN(x)/dm = (\rho(x)/M)\delta(M - m)$. There is a simple relation giving the

Table 2: The optical depth ($\tau$), the event rate ($\Gamma$), and the average event duration ($\langle T \rangle$), for microlensing events in Baade's window. The results for disk DM and faint disk stars are taken from Ref. [15].

| Model | $\tau$ | $\langle x \rangle$ [kpc] | $\Gamma$ [$\frac{\text{events}}{\text{yr }10^6\text{stars}}$] | $\langle T \rangle$ [days] |
|---|---|---|---|---|
| Halo | $2.6 \times 10^{-7}$ | 5.3 | $10/\sqrt{m_{0.05}}$ | $6\sqrt{m_{0.05}}$ |
| Spheroid | $5.1 \times 10^{-7}$ | 7.1 | $25/\sqrt{m_{0.05}}$ | $5\sqrt{m_{0.05}}$ |
| Faint stars | $(2.9 - 9.6) \times 10^{-7}$ | 4.5 | 2.2–7.5 | 30 |

average event duration $\langle T \rangle = (2/\pi)\tau/\Gamma$, whose values are shown in the last column of Table 1. Note that the EROS/MACHO events, with durations $T = 17, 9, 14, 27$ and $30$ days suggest lens masses in the brown dwarf range of $m \sim 10^{-2}$–$10^{-1} M_\odot$ for both models. At the time of this writing, both the EROS and MACHO experiments have accumulated a statistics of approximately $10^7$ stars yr and have seen two and three events respectively. We then see from Table 1 that, unless the (still unavailable) observational efficiencies were below 10%, the spheroid predictions are quite consistent with observations while the halo ones seem to be somewhat in excess.

A third microlensing experiment, the OGLE collaboration [13], as well as the MACHO team, have been looking for microlensing events on stars in the galactic bulge. Here the contribution of the spheroid dark objects is expected to be particularly important, due to the rise in the spheroid density towards the galactic centre. However, a strong background for these observations is due to the faint stars in the disk itself [14–15], which can be computed from the knowledge of the disk stellar mass function [16] and taking into account the star rotation. It is also important in this case to take into account the large velocity dispersion of source stars $\sigma^{Bulge} \simeq 105$ km/s. This is performed by integrating Eq. (9) over the source velocities using a Maxwellian distribution. We show in Table 2 the same quantities as in Table 1, but for observations in Baade's window (BW), for the spheroid, the halo and for disk stars.

Figure 2 shows $(L/\tau)d\tau/dx$ for observations in the BW. We see that in this case the main contribution to the spheroid rates comes from the inner 1–2 kpc around the galactic centre. The values of $\Gamma$ reported in Table 2 indicate that the expected rates from the spheroid are larger than those from the halo and that the main characteristic of these events is their short duration, of just a few days for brown dwarf masses. Of course, to relate the 'theoretical' rates with the actual observations, a knowledge of the experimental efficiencies is required. These are however not yet available, but since at present typically one image per field is taken each night, not all nights are good for observation, and the selection criteria for the events require typically five points in the light curve above the amplification threshold, the efficiency becomes very poor for events of durations shorter than a week, where most of the spheroid (and halo) events are expected.

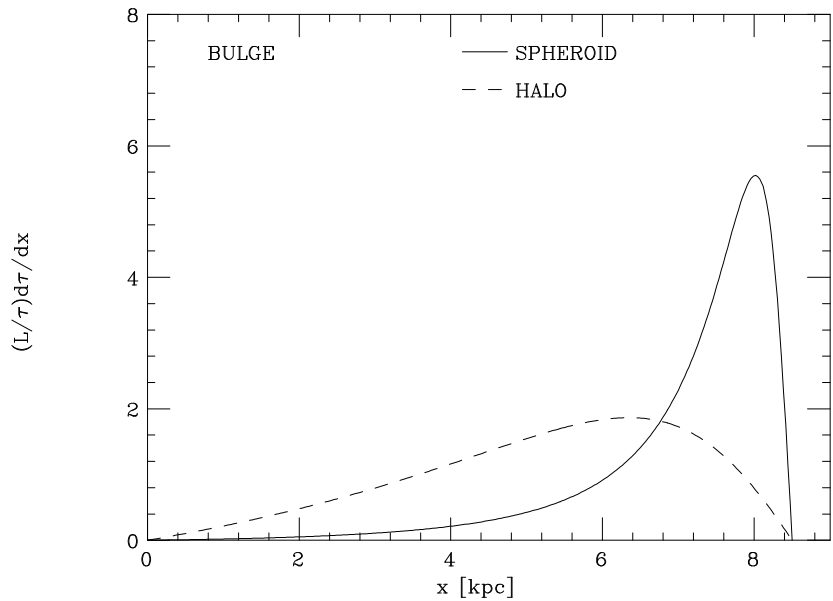

Fig. 2: Contribution to $\tau$ vs. the distance to the lensing objects $x$ for the Bulge.

Figure 3 shows the theoretical rates for events with duration $T > T_0$ as a function of $T_0$, i.e. $\Gamma(T > T_0) = \int_{T_0}^{\infty} dT\, (d\Gamma/dT)$. We can see that they are very sensitive to the minimum detectable duration $T_0$. Hence, it seems crucial for bulge observations to reach good sensitivity to events of a few days by increasing the number of exposures to the same field during the same night. Another attractive strategy would be to combine observations of the OGLE and MACHO experiments of the same stars, since if this were possible the different terrestrial longitudes of the two observatories (Chile and Australia, respectively) would provide a more complete time coverage of events of short duration. It should be noted that the events observed up to now by OGLE, with durations of 11, 12, 14, 26 and 45 days, and those of MACHO, with durations of 10, 21, 24 and 25 days, are probably due mainly to faint disk stars.

Finally, as discussed in Ref. [6], there are other signatures that could allow to distinguish among halo and spheroid lenses, such as the angular dependence of the bulge rates (using measurements in fields other than the BW) or the angular dependence of the rates in the Andromeda galaxy, if those observations become feasible.

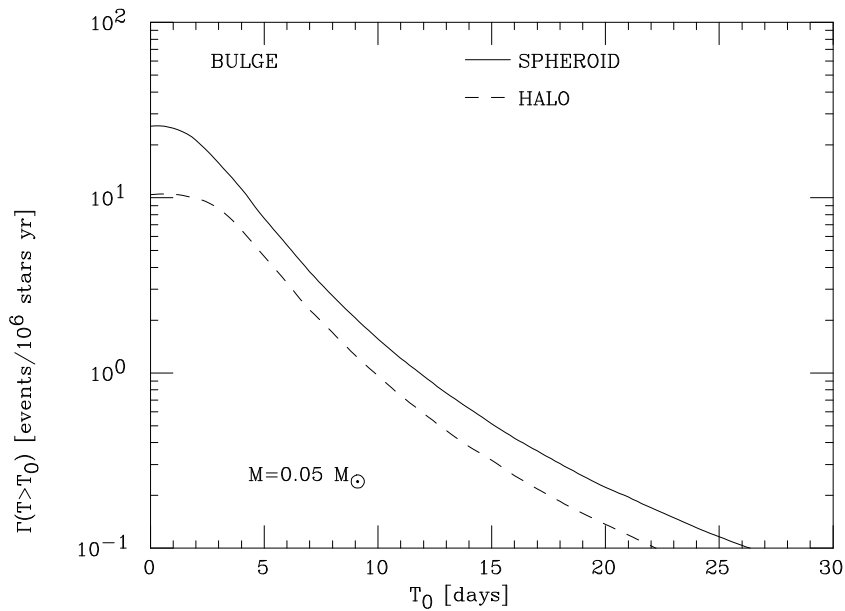

Fig. 3: Rate $\Gamma(T > T_0)$ vs. $T_0$ for bulge observations.

The explanation of the microlensing events as due to a spheroid dark population is particularly attractive in the light of its implications for cosmology and structure formation. A recent determination of the 'primordial' deuterium abundance [17], combined with the nucleosynthesis predictions, suggests that the baryonic content of the Universe is just of the order of the amount of luminous matter. This does not leave much room for large quantities of dark baryons, as would be required if galactic haloes were baryonic, but is perfectly consistent with 'mostly dark' baryonic spheroids. If non-baryonic dark matter also exists, as seems necessary to make compatible the determinations of $\Omega$ at large scales with the nucleosynthesis predictions, there would be no reason for it not to fall in the potential wells of the galaxies and to be around us. The haloes would then be the natural place for it to reside. The presence of non-baryonic dark matter (of the so-called 'cold' type) in the galaxies is also desirable in the light of our present understanding of structure formation in the Universe, in which baryons fall in the potential wells created by the cold dark matter density perturbations. Due to the dissipative nature of baryons, we expect them to become more concentrated towards the centre of the galaxy than the non-dissipative dark matter after the gravitational collapse, and this would explain the different spatial distribution of the two components.